\documentclass[preprint]{aastex}

\shorttitle{Effectiveness of the CFHT}
\shortauthors{Crabtree and Bryson}

\begin{document}


\title{The Effectiveness of the Canada-France-Hawaii Telescope}


\author{Dennis R. Crabtree}
\affil{Canadian Gemini Office, Herzberg Institute of Astrophysics,\\ 
National Research Council Canada, Victoria, BC V9E 2E7, Canada}
\email{Dennis.Crabtree@nrc.ca}

\author{Elizabeth P. Bryson}
\affil{Canada-France-Hawaii Telescope, Kamuela, HI 96743}
\email{bryson@cfht.hawaii.edu}

\begin{abstract}
We have investigated the productivity and impact of the Canada-France-Hawaii
Telescope (CFHT) during its twenty-year history. CFHT has
maintained a database of refereed publications based on data obtained with CFHT since
first light in 1979. For each paper, we analysed the cumulative number of citations 
and the citation counts for each year, from data supplied by the NASA Astrophysics 
Data System (ADS).  We have compared citation counts retrieved
from the ADS with those from the Institute for Scientific Information (ISI) for a small 
sample of papers. We have developed a procedure that allows us to
compare citation counts between older and newer papers in order to judge their
relative impact. We looked at
the number of papers and citations not only by year, but also by the instrument used to
obtain the data. We also provide a preliminary look as to whether programs given
a higher ranking by the Time Allocation Committee (TAC) produced papers with a higher
number of citations.
\end{abstract}



\section{Introduction}

Astronomers use telescopes to investigate a wide range of scientific problems
with almost as diverse a range of instrumentation. The output from these
investigations provides an immense amount of data that needs to be reduced and
analyzed. The results from the investigation are published in scientific
journals, and these papers have an impact, small or large, on future
observational and/or theoretical investigations.

Two measures of the effectiveness of a telescope are the number of papers
published in refereed journals that are based on data obtained by the
telescope, and the citation count of those papers.  The effectiveness, or lack
therein, of a telescope can have far-reaching consequences. For example, in
Canada the effectiveness of a single major telescope, the Canada-France-Hawaii
Telescope (CFHT), may have a significant impact on the funding of future
telescopes. \citet{abt85} compared the impact of two facility telescopes (the CTIO
and KPNO 4-m) with that of two telescopes run by private observatories (the
Lick 3-m and the Palomar 5-m). He found no significant difference.
\citet{tri95} compared the impact of large US
optical telescopes for papers published in 1990-1991. More recently \citet{ben01}
compared the scientific impacts of telescopes world-wide based
on their contributions to the 1000 most-cited papers (1991-98) and the number
of papers published in Nature between 1989-1998. They found that CFHT was the most
productive and most highly-cited of all 4-m class telescopes during this time
period.

Productivity, as measured by the number of papers, and impact, as measured by
citation numbers are the two measures we will use to assess the effectiveness
of the CFHT over its approximately twenty-year history. Simply counting the
number of papers in refereed journals is an easy way to measure effectiveness
but misses completely the influence these papers have on the field. It should
be noted that citation numbers are not a perfect measure of a paper's impact, nor are
they necessarily a measure of the paper's scientific value. In this
contribution we will examine the productivity and impact history of CFHT papers. We
will also examine the productivity and impact of the various instruments that
have been used at CFHT during its twenty years of operation. Finally, we will look at
how the citation counts for published papers are related to the grade assigned
the original observing proposal by the Time Allocation Committee.

\section{The Data}

CFHT maintains a database of publications in refereed journals that are
based on data obtained with the telescope. The database contains
information on 1065 papers published between 1980-1999. Papers are identified
from four main sources: reprints submitted by authors, scanning of all major
journals, observers' time request forms, and searching NASA's Astrophysics Data
System (ADS) for papers referring to CFHT in the abstract.
The following criteria are used to judge whether a paper is considered a CFHT
publication:
\begin{quote}
"A paper must report new results based on significant observational data
obtained at CFHT or be based on archival data retrieved from the CFHT archive.
If data from multiple telescopes are included, the CFHT data should represent a
significant fraction of the total data".
\end{quote}
A staff astronomer examines each paper to judge whether it meets these
criteria. Although an author may footnote a paper to indicate that it is based on
CFHT observations, the paper  may not meet the criteria for inclusion in the database.
In our view, this rigorous emphasis on validation of all papers by astronomers
within CFHT makes the database unique. 

The CFHT publication information is maintained within a Microsoft Access
database. Several routines, written in Visual Basic for Applications within the
database, query the ADS for information on each publication. These routines
utilize an Internet Data Transfer Library \citep{ash98}
downloaded from the Internet.
The software generates the appropriate query as a URL, sends the URL to the ADS,
and parses the returned text to extract the relevant information.
The information for each publication in the ADS is accessed by a publication
bibliography code, bibcode, which is generated from the year, journal, volume
and page information for a publication. One of the many services ADS provides
is a verification utility that returns a yes/no as to whether a particular
bibcode is valid. For each entry in our database, the ADS bibcode is generated
from the publication information and verified with the ADS. The information for
those entries with invalid bibcodes is checked and updated, then a new bibcode
is generated and verified. This verification of each paper's bibcode ensures
that we have the correct publication information for each entry. Once each
publication has a valid bibcode, the ADS is queried for the full title, list of
authors, the number of citations, and the number of self-citations (ones in
which the first author of the cited and citing paper are the same person).
Finally, the bibcodes of each citing paper and the number of citations by year
of the citing paper are recorded for each publication.
The instrument, or instruments, used to acquire the data used for each
publication was identified by browsing each of the papers.

This use of the ADS allows us to verify the basic bibliographic information,
obtain a complete list of authors, and collect the citation data for each
publication. The citation information in the ADS is incomplete \citep{kur00}
. Much of the citation information in the ADS is based upon a subset of
the Science Citation Index purchased from the Institute for Scientific
Information (ISI) by the ADS. This subset is seriously incomplete in referring
to articles in the non-astronomical literature, as it only contains references
that were in the ADS when the subset was purchased. The ADS currently builds
citation links itself for all publications in its database. The ADS does not
include many physics journals but does include a subset of conference
proceedings.

\subsection{Comparison of ISI and ADS Citation Counts}

ISI, an established and reputable commercial firm, has been considered the
best resource for citation information among astronomers and librarians
for many years. However, the ADS provides publication and citation information
from the Web at no cost.  How does the citation information obtained from these
two sources compare?
We selected three highly cited CFHT papers: \citet{car96}; \citet{cow96};
\citet{lil96}, and performed a detailed analysis of citations to
these papers using both the ADS and ISI (through the online service DialogWeb). 
While the
total number of citations to the three papers from ISI/ADS are remarkably
similar (146/153, 125/124, 172/177), there are interesting differences in the
details of the citing papers. The number of citing papers in common to ADS and
ISI for the three papers is 142, 109 and 165. Each database missed several
citing papers that the other one included. ISI tended to find citations from
physics journals missed by ADS, while ADS had some conference citations and
citations from the major journals that were missed by ISI. The citing papers in
the major journals  were missed by
ISI primarily due to incorrect citations (e.g. wrong year or volume) in the
citing papers. Our conclusion from this detailed look at a small number of
papers is that, on average, the ADS provides citation numbers that are consistent
with those obtained from ISI and any differences will have a minimal impact on
our study.

\section{CFHT's Productivity and Impact}

We define two terms that we will use throughout the rest of the paper:
productivity and impact. Productivity refers to the number of publications in
the context of the telescope, an instrument or a particular researcher.
Productivity is not the same as scientific impact. Impact is usually measured
by using citation numbers. Overall impact is measured by summing citation
numbers of all the relevant papers. The average number of citations per paper
(CPP) measures the average impact.

\subsection{Citation Histories}

For each entry in the CFHT database, we have retrieved the year of every citing
paper and stored the total number of citations for each year in the database. A
paper published in 1990, for example, has the number of citations received for
each year from 1990 to 1999. These data allow us to investigate the citation
rate as a function of the number of years since publication.
The solid curve in Figure 1 shows the average citations per paper (CPP)
as a function of the number of years after publication for all papers in the
database with citations. There are some citations in the year of publication
for papers published early in the year; for example, a paper published in January may
receive a citation in November. As the number of years since
publication increases, the number of papers included decreases since the
relevant data for all papers doesn't yet exist. The papers published in 1999
are included in only the data point for zero years after publication, and 1998
papers are included in the zero and one year data points, etc. This curve peaks
at two years after publication and has a fairly smooth decay after that. It has
been known for many years that the number of citations a paper receives
declines exponentially with the age of the paper (e.g. \citet{bur60}).
This is true of CFHT publications as well. The dashed line in Figure 1 is the
fit of a simple exponential decline in citations with a half-life of 4.93 years
beginning two years after publication.

Our analysis does not include a correction for a growth in publication numbers
over the period 1982-1999. \citet{abt81} found a half-life of around twenty years for
papers published in the 1961 issues of \apj, \apjs\ and \aj. However, he pointed 
out that the growth in the number of papers published over the eighteen year-period he
gathered citation numbers was part of the reason the half-life was so long.
\citet{pet88} shows that number of papers published in those three journals
increased by 4.6 times over this period. Kurtz et al. (2000) show that the
number of papers in their Big8 journals (\apj, ApJL, \apjs, \aap, \aaps, \mnras, 
\aj\ and \pasp) increased at approximately a 3.7\% yearly rate between 1976 and 1998.
The astronomical literature has doubled between 1982 and 1999, the years
covered in our study.

Has the electronic distribution of preprints and journal articles changed the
citation history of papers? To examine this question, we divided the CFHT papers
in two groups: those published between 1984 and 1991, and those published
between 1992 and 1999. There were 428 and 522 papers in these two groups. In Figure
2 the average citation rate for these two periods is shown along with the fit of a simple
exponential model for each period. The citation rate for the newer papers
clearly declines more rapidly than that of the older papers. The half-life of
the older papers is 7.11 years while the half-life for the newer papers is 2.77
years.  We believe that if one were able to sample citation rates monthly, 
the citations for an average paper in the more recent dataset would peak less 
than two years after publication.
This is the result of more rapid dissemination of results by the
electronic distribution of pre-prints (astro-ph) and journal articles.
The faster decline in citations for the recent subset also indicate that new
results supersede earlier results more quickly than in the past.

\subsection{Comparing Papers from Differing Years: A Standard Citation Measure}

Comparing the citation numbers for papers published in different years is
difficult since the number of citations to a paper increases with time. We have
established a method for estimating the total number of citations that a paper
can be expected to achieve after a suitably long period. How do we compare
papers published over almost twenty years, given the natural growth in citations
with time? We have used the average citation history of all CFHT papers to
define a growth curve for citations (Figure 3). This curve shows the percentage
of the final number of citations, defined as the number eighteen years after
publication, for an average paper versus the years since publication. Using
this curve we can estimate the final citation count (FCC) for each paper given
a citation count and the number of years since publication.

\section{Productivity and Impact History}

The first CFHT paper was submitted in May 1980 and was published in
August of that year \citep{vdb80}. CFHT's productivity (Figure 4) rose
more or less continuously through the 1980s until it reached a fairly
constant level of around seventy-five papers per year between 1991
to 1997. It took approximately ten years for CFHT to hit its stride
and reach a consistently high level of paper production. A telescope's
productivity in any one year is linked to many factors such as weather,
competitiveness of the available instruments, and the reliability of
instruments and the telescope, all in the several years before the
year of publication. We attribute the increase in publications during
the first ten years of CFHT to the increase in the reliability of
both the instruments and the telescope and to the development of more
competitive instruments. There are two possible reasons the number of
CFHT publications may be in a slow decline. First, as more 8-10 meter
telescopes come on-line, CFHT is no longer a forefront facility. Second,
the use of large mosaic CCD cameras has increased at CFHT. These generate
a tremendous amount of data, and the time from acquisition of data to
the publication of results has likely increased.

Trimble (1995) studied the productivity of large, American optical
telescopes including CFHT. She compiled publication data for an eighteen
month period beginning January 1990, by examining the major North American
journals: \apj, ApJL, \apjs, \aj, \pasp.  According to Trimble's list,
CFHT ranked fourth in productivity behind the CTIO 4-meter, Palomar and
the KPNO 4-meter; and, as Trimble notes, many CFHT publications appear
in journals not included in her study.  Taking all of the 1990 papers
and half of the 1991 papers, we count sixty-seven CFHT papers (Trimble
counted 58.6) that were published in the major North American journals
during this period.  (Trimble pro-rated each paper based upon the number
of telescopes used in the paper, which we have not done.)  Our database
contains one hundred one CFHT papers published in {\it all} refereed
journals during this period. If we correct this number by the same factor that
our earlier number differs from Trimble's for only North American journals, we
end up with a total of 88.3 papers. The total number of CFHT publication
changed significantly by including publications from all journals. While
the other telescopes undoubtedly had publications in non-North American
journals, except for the Anglo-Australian Telescope, their numbers would
not have increased as significantly.  Thus, any future study of papers
and citations, especially those that compare different facilities, should
include all major journals.  The average CPP for all papers in a given
year, by year of publication, is shown in Figure 5. One would expect the
average CPP to grow smoothly with time since publication. However, due
to the relatively small number of papers in any given year, the average
CPP can be influenced by a small number of highly cited papers. For
example, the bump in 1996 is due to two highly cited papers (Lilly et
al. 1996, Carlberg et al. 1996) that are based on data taken with MOS,
the Multi-Object Spectrograph. The fluctuations in citation numbers are
much higher in earlier years when the number of papers was smaller.

\subsection{Publications and Citations by Journal}

Most CFHT observers are from Canada, France or the University of Hawaii (UH).
The French tend to publish in European journals, mainly \aap, while Canadian and
UH researchers favor North American journals. How are CFHT publications
distributed across the major journals? The distribution of publications across
eight journals (we include ApJL with \apj) is shown on the left side of
Table 1. In addition, each paper has been tagged as belonging to one of the
three partners based upon the affiliation of the first author or the agency
that granted time for the observations. (Canada grants some time to
international researchers). The majority of CFHT papers have been published in
the three major journals - \apj, \aap\ and \aj\ account for more than 78\% of CFHT
papers.

ApJ has the most publications with 33\% of all CFHT publications, while \aap\ receives
25.8\% of the publications. The breakdown of publications by journal for
different years shows an interesting change. In 1996/1997 33\% and 25\% of papers
were published in \aj\ and \aap\ respectively, while for 1998/1999 the numbers
were 25\% and 45\%.  One explanation for this change is that the French are
publishing more and the Canadians/UH, less in recent years. 
Only 75 (18.9\%) of the French papers appeared in American journals while only
53 (9.7\%) Canadian papers, and 6 UH papers (5.1\%), appeared in non-North
American journals. There is a very strong trend for European authors to publish
in European journals and North American authors to publish in North American
journals. This may be a result of the fact that \aap\ has no page charges and
subsequently the French do not have a large budget for page charges. This tendency for
authors to publish on their side of the ``Atlantic Ocean'' is particularly
meaningful for any comparison of publication activity levels between North
American (only) and multinational observatories.
The distribution of citations per paper (CPP) across the journals is shown in
Table 2. The three major journals, \apj\ (including the Letters), \aap\ and \aj,
account for 84.7\% of the citations to CFHT papers. While ApJ papers account
for 33\% of the CFHT total, these papers received almost half  (48.1\%) of the
citations. \aap, \aaps, \aj, \mnras\ and \pasp\ all have a citation rate lower than
the average CPP of 20.35. Nature has the highest CPP of any journal (2.8\%);
and yet these represent only 1.6\% of CFHT papers.

\subsection{Publications and Final Citation Count by Instrument}

The primary instrument used to acquire the data was identified for each
publication. In a few cases, several instruments were grouped together into a
single category. For example, FP refers to several different ``Fabry-Perot''
instruments, Coud\'{e} refers to the two Coud\'{e} spectrographs that have been used at
CFHT, and ``Direct Imaging'' combines several different direct imaging cameras
that have been used at CFHT over the years. HRCam \citep{mcl89}, which
incorporated fast tip-tilt correction, and, MOCAM \citep{jcc96} and UH8K
\citep{met95} two mosaic cameras are identified separately from other direct
cameras because they represent new technologies, and we wish to track their impact
directly. A total of 39 distinct instrument or instrument categories were
identified; however, a large number of instruments produced a very few papers
and approximately 70\% of CFHT papers were produced by the top five
instrument/instrument categories.
Table 2 shows the number of papers, the FCC per paper and the FCC per night of
scheduled telescope time for the top-ten paper producing instruments. The
number of nights of scheduled telescope time was determined by looking at each
semester's schedule from 1982 onward and counting the number of nights for each
instrument/instrument group. CFHT is known for its exceptional image quality,
and it is no surprise that direct imaging has produced the highest number of
papers. It also has the highest efficiency of turning scheduled nights into
citations. The two Coud\'{e} spectrographs and the Multi-Object Spectrograph (MOS)
produced the 2nd and 3rd highest number of papers. The CFRS \citep{lil96} and
CNOC \citep{car96} studies are a large contributing factor to the high impact of
MOS.

\section{CFHT's Prolific Authors}

Who have been the most prolific authors over the twenty years of CFHT publications?
Table 3 shows the top nine most prolific authors with the total number of
publications, the number of publications in four of the major journals, the
total number of citations to their papers, their average CPP and their
projected FCC assuming they publish no more papers based on CFHT data. The most
prolific authors have favoured North American journals. The two French authors
in this list have 45\% of their papers in North American journals as
compared to only 18.9 \% of all papers designated as French. Citations will be
discussed more thoroughly in the next section. However, it is clear that the
average CPP for these authors varies significantly.

\subsection{Self Citations}

The issue of self-citations (ones where the first author of cited and citing
paper are the same) is one frequently asked of (and discussed by) librarians.
What is the average self-citation rate? As Trimble points out, this number is
difficult to determine exactly. Authors do not always use a consistent name
(first name or first initial, for example) which may lead to the incorrect
counting of self-citations. We have counted self-citations for CFHT papers by
matching first authors on the cited and citing papers. The average
self-citation rate for all CFHT papers is 6.3\%. On average, corrections to
citation numbers for self-citations are not important. However, the
self-citation rate for individual papers can be much higher. There are almost
ninety papers with a self-citation rate of 30\% or more and many of these have ten or
more citations. Also, certain authors tend to favour their own work. Several
authors with four papers, or more, have average self-citation rates of 20\% or
higher. Highly cited papers had much lower than average self-citation rates.
The twenty most cited papers in the database have an average self-citation rate
of 3\%, less than half of the average rate for all papers.

\section{CFHT's Most Highly Cited Papers}

We computed the final citation count (FCC) for each CFHT paper in the database
using the growth curve described above. The ten papers with the highest FCC are
listed in Table 4. It is interesting to note that the top three papers in this
list have the same first author, Simon Lilly. Lilly also has the highest total
FCC, summed over all papers, of any author in the CFHT database. Two of the top
three papers are based on data from a large, ambitious project undertaken with
a new forefront instrument. The top three papers all have the word ``survey'' in
their title as well.

\section{How Effective is the Time Allocation Committee?}

The observing time requested by proposals to use any large telescope
such as the CFHT, generally outnumbers the available time by a significan t
factor. A Time Allocation Committee (TAC) is established to review and
to rank the submitted proposals, and, in classical scheduling, only the
highest-ranked proposals make it to the telescope. However, in queue
scheduling, the relative ranking of the proposals will be an important
factor in determining which programs are actually executed. The role of
the TAC becomes even more critical in the era of queue scheduling. How
effective is the TAC in judging the scientific merit of proposals? Most
would agree that almost all programs that reach the telescope will
likely produce a scientific publication if the weather and the equipment
co-operate. However, one would expect that the more highly ranked
proposals will, on average, produce publications with a higher impact,
i.e. number of citations. We feel this evaluation of the TAC process is
important as several large telescopes undertake queue scheduling.

Observing time at CFHT is allocated by country: 42.5\% for each of Canada and
France, and 15\% for the University of Hawaii. Each country runs its own TAC,
which assigns the grades. The International TAC meets to deal with scheduling
conflicts and program overlaps.
We have identified the original proposal associated with twenty-two CFHT papers
published between 1997 and 1999. One of us (D.C.) has access to the TAC ranking
for these proposals as he served as Senior Resident Astronomer for three years.
We are thus able to look at the correlation between the TAC ranking of the
proposal and the predicted FCC for the papers resulting from those
observations. We selected only those papers that were based on a proposal that
used only data from CFHT and data from a single observing run. Figure 5 shows
the FCC versus TAC ranking (a small number is a higher ranking) for these
twenty-two
proposals. Except for one highly ranked, highly cited study, there is a weak
inverse correlation from a simple linear fit to the data. Another
interpretation of the data is that highly-ranked studies ($<$ 0.4) show small
scatter around a constant number of citations, while lower-ranked studies show
a much larger scatter. Is the TAC being conservative and ranking sure bets
higher while riskier studies end up with lower rankings? We want to emphasise
that this result is very preliminary, as only twenty-two papers are included, and it
relies on the predicted final citation counts. We found no dependence of the
FCC on the number of nights of telescope time awarded for the same twenty-two
programs.

\section{Conclusions}

We have studied the productivity and impact of the CFHT over its twenty-year
history by looking at the number of papers in refereed journals and the number
of citations to these papers. It took ten years for CFHT to achieve and maintain
a high level of paper production. We attribute this to a fairly long
commissioning period for the telescope and the time to develop a competitive
suite of instrumentation. Direct imagers (photographic plates, CCD imagers)
have been CFHT's most productive instruments, both in the number of papers and
the number of papers per night of scheduled telescope time. The excellent image
quality at CFHT is a significant factor in direct imaging's high productivity.

We retrieved citation counts and the years of the citing papers from the ADS for 
all CFHT papers in our database. Using this data, we developed a procedure for 
estimating the number of citations that a paper can be expected to receive after a 
period of almost twenty years. This estimation allowed us to compare the citation numbers 
for papers from different years and to compare the impact of different instruments.  
The instrument that produced the papers with the highest impact (average
citations/paper) was the Multi-Object Spectrograph (MOS), which was used in the
highly cited CFRS and CNOC studies. Direct imaging had the second highest
impact. In looking at the number of citations/night of allocated time, direct 
imaging had the highest impact, followed by MOS and the two Coud\'{e}
spectrographs. The efficiency of converting observing nights into papers or
citations varies
considerably between instruments. For example, there is a factor of five difference
in the average final citation count per night between ``Direct Imaging'' and
the FTS. In order to maximize a telescope's impact, one might consider offering
only the ``high-efficiency'' instruments.

Finally, a look at the correlation between the predicted final citation count
and the TAC ranking of the observing proposal, showed a weak negative
correlation, i.e. lower-ranked proposals end up with a higher number of
citations. An alternative interpretation has higher-ranked proposals with a
lower number of citations with a small scatter. Lower-ranked proposals have
more scatter in the number of citations and some of these end up with
significantly more citations than most of the higher ranked proposals.

\acknowledgments

We acknowledge the Canada-France-Hawaii Telescope and the Herzberg Institute of
Astrophysics for their support of this project. We thank Pierre Couturier
for his impetus in initiating the analysis of CFHT publications. We 
also thank Gordon W. Bryson for his editing and Virginia Smith for her assistance during her
mentorship at CFHT. This research has made use of NASA's Astrophysics Data System
Bibliographic Services. 





\clearpage






\begin{figure}
\figurenum{1}
\epsscale{0.8}
\plotone{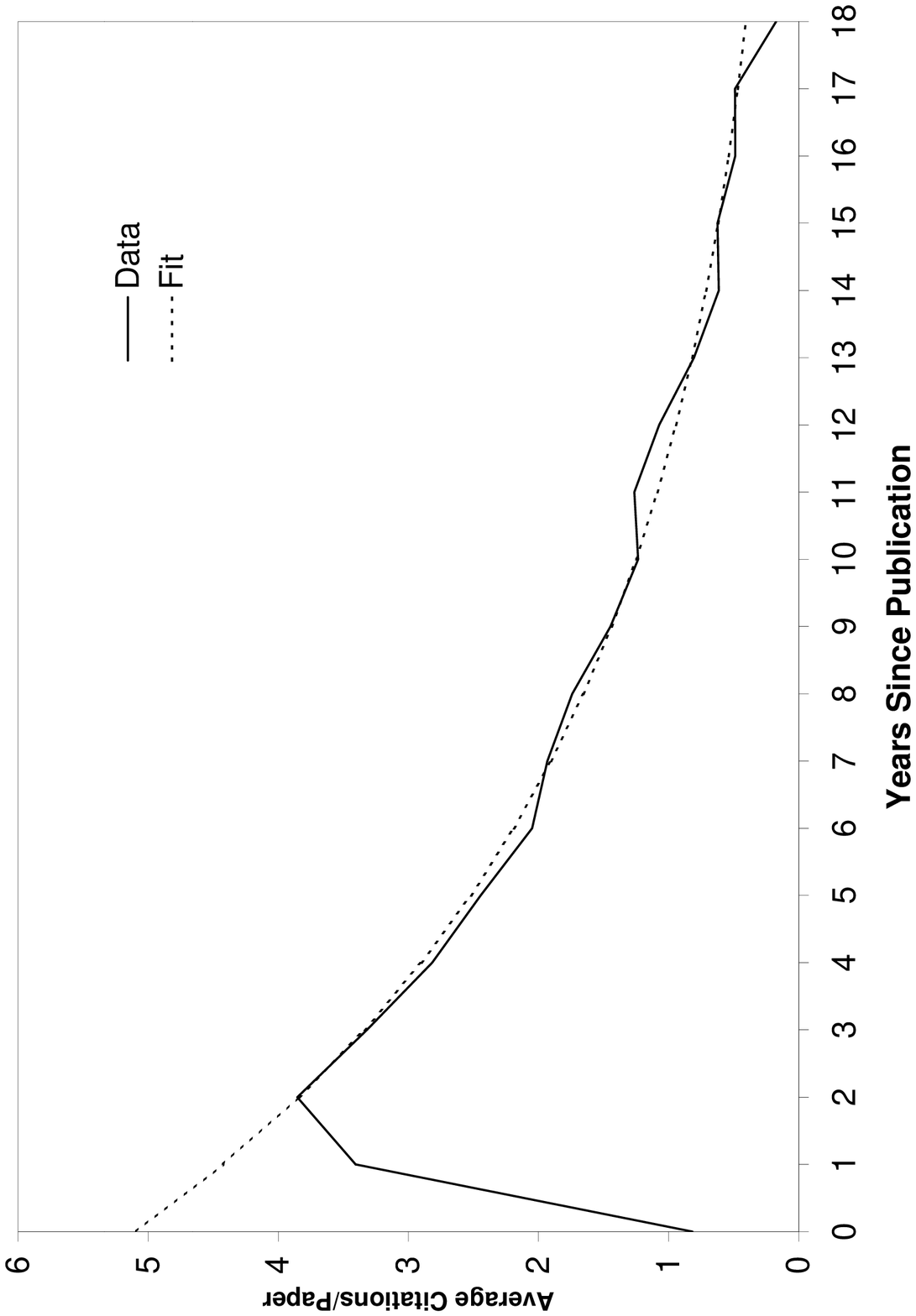}
\caption{The average citation rate as a function of the number
of years since publication. The solid line is the data for all CFHT papers and
the dashed line is the fit of a simple exponential decline with a {\it
half-life} of 4.93 years.}
\end{figure}

\clearpage 

\begin{figure}
\figurenum{2}
\epsscale{0.8}
\plotone{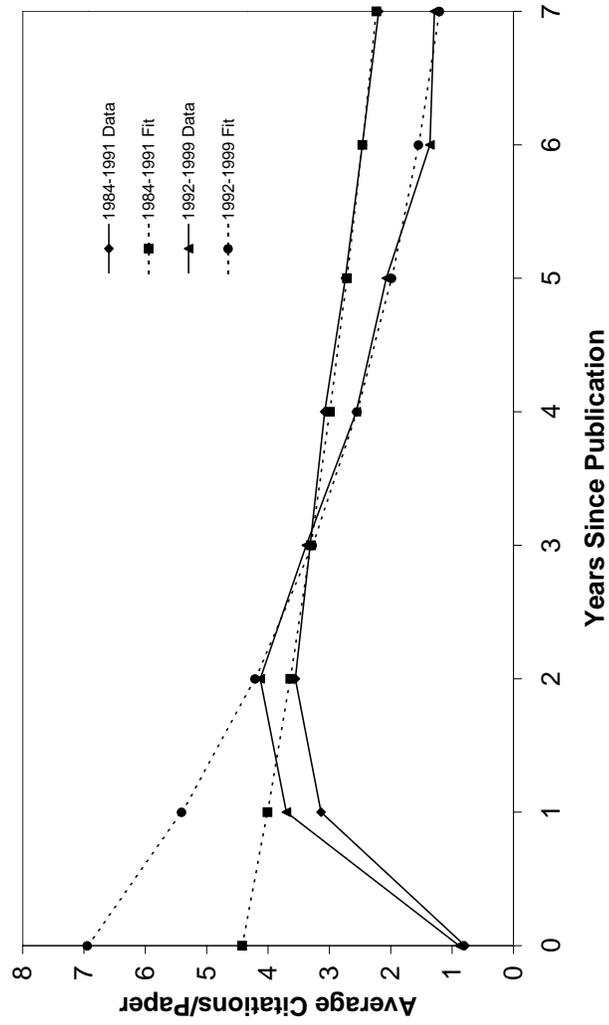}
\caption{The average citation rate as a function of years since publication
for CFHT split into two groups; 1984-1991 papers and 1992-1999 papers.}
\end{figure}

\clearpage 

\begin{figure}
\figurenum{3}
\epsscale{0.8}
\plotone{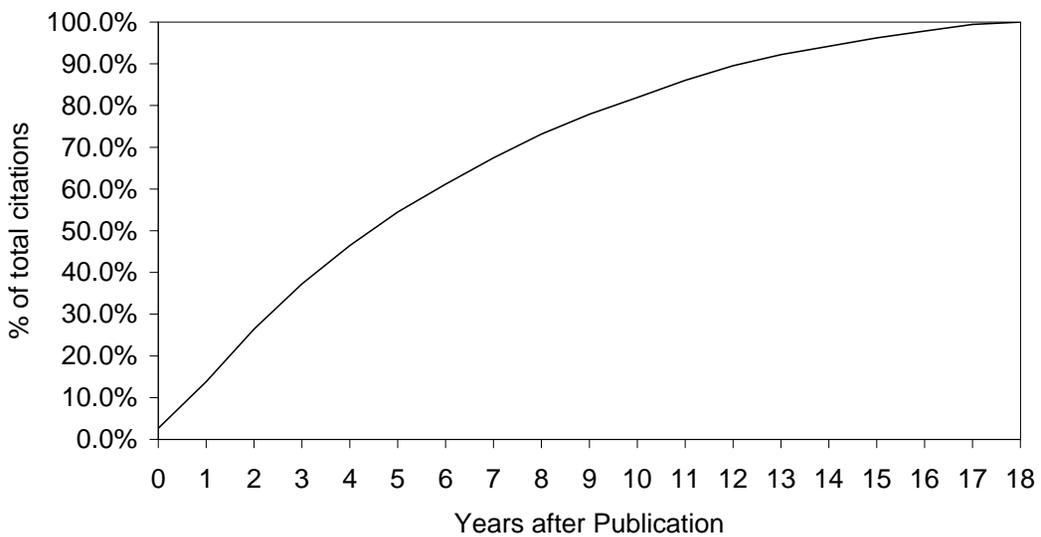}
\caption{Citation count growth curve for all CFHT publications.}
\end{figure}

\clearpage 

\begin{figure}
\figurenum{4}
\epsscale{0.8}
\plotone{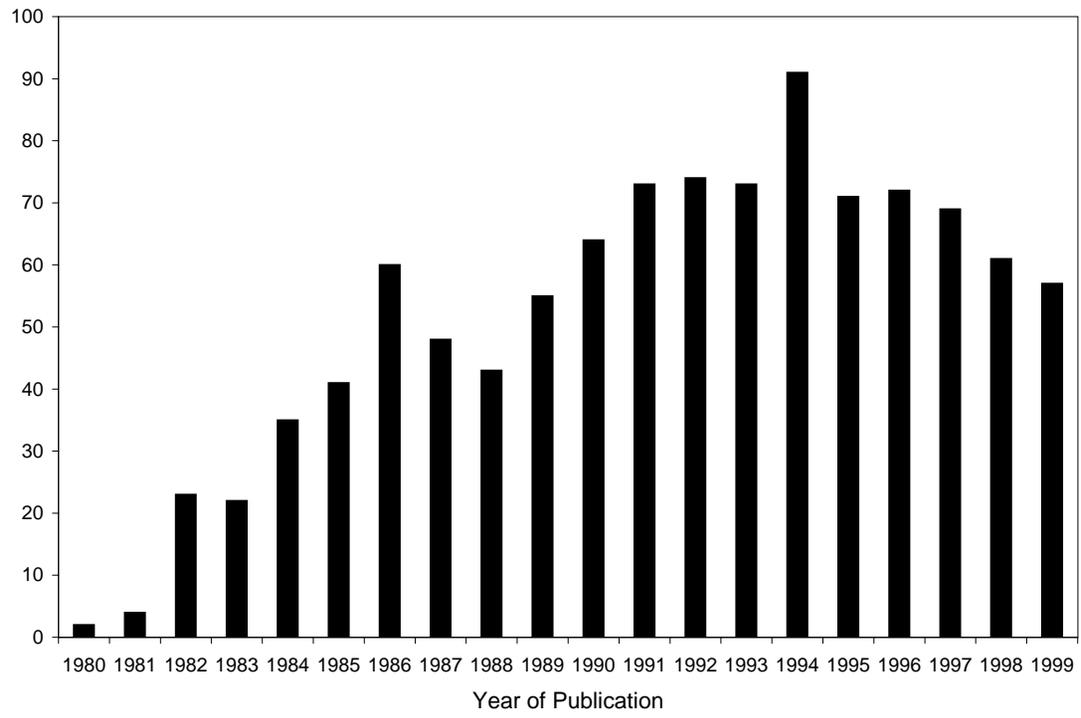}
\caption{Number of CFHT Papers by Year of Publication}
\end{figure}

\clearpage 

\begin{figure}
\figurenum{5}
\epsscale{0.8}
\plotone{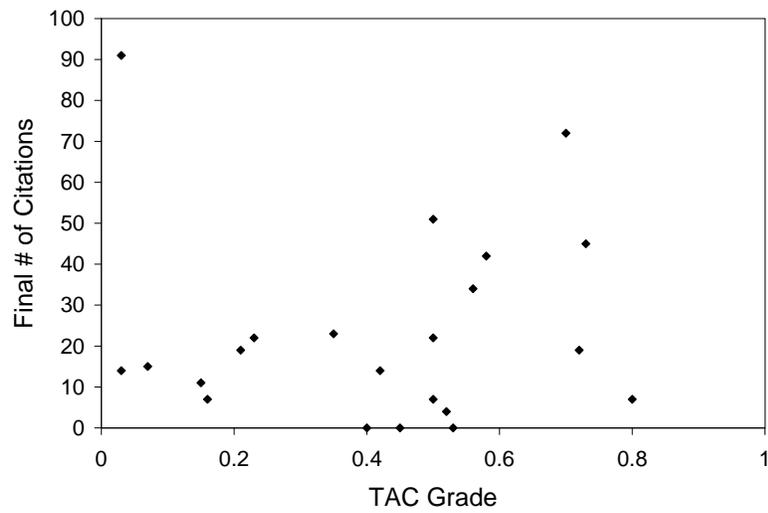}
\caption{The predicted {\it final} number of citations versus the Time Allocation
Committee (TAC) grade assign the original proposal for a subset of CFHT 
publications. A lower TAC grade is ``better''. The highest ranked
proposal will have a TAC grade near 0.0, while proposals that just make it on
the schedule have a grade of 1.0. Note the single point with a very high TAC grade
and a predicted high number of citations.}
\end{figure}

\clearpage

\begin{deluxetable}{lccccccccc}
\tablecaption{Distribution of Papers and Citations by Journal. \label{tbl-1}}
\tablecolumns{10}
\tablehead{
\colhead{} & \multicolumn{3}{c}{Number of Papers} & \colhead{} & \colhead{} & \multicolumn{4}{c}{Citations/Paper}\\
\cline{2-5} \cline{7-10}
\colhead{Journal} & \colhead{All} & \colhead{C} & \colhead{F} & \colhead{H} & \colhead{} & \colhead{All} & \colhead{C} & \colhead{F} & \colhead{H}}
\startdata
\aap & 275 & 34 & 239 & 2 & & 14.86 & 8.79 & 15.82 & 4.00\\
\aaps & 37 & 4 & 32 & 1 & & 11.92 & 4.25 & 12.91 & 11.00\\
\aj & 209 & 174 & 10 & 25 & & 18.70 & 18.06 & 15.00 & 24.64\\
\apj & 351 & 223 & 59 & 69 & & 29.73 & 29.20 & 22.34 & 37.77\\
\apjs & 22 & 18 & 1 & 3 & & 28.00 & 30.28 & 2.00 & 23.00\\
\mnras & 20 & 10 & 10 & 0 & & 9.25 & 11.90 & 6.60 & 0.00\\
\nat & 18 & 5 & 9 & 3 & & 33.72 & 47.20 & 19.89 & 64.00\\
\pasp & 82 & 66 & 5 & 11 & & 11.02 & 11.33 & 7.60 & 10.73\\
\tableline\tableline
Total/Average & 1065 & 549 & 397 & 117 & & 20.35 & 21.42 & 15.75 & 31.30\\
\enddata
\end{deluxetable}

\clearpage

\begin{deluxetable}{p{6.0cm}cccc}
\rotate
\tablecaption{Papers and Citations by Instrument. \label{tbl-2}}
\tablehead{
\colhead{Instrument} & \colhead{Papers} & \colhead{Papers/night} & \colhead{FCC/paper} & \colhead{FCC/Night}}
\startdata
Direct Imaging & 358 & 0.36 & 35.47 & 12.86\\
Coud\'{e} Spectrograph & 169 & 0.25 & 28.37 & 6.97\\
Multi-Object Spectrograph & 75 & 0.19 & 48.38 & 9.92\\
Fourier Transform Spectrometer & 64 & 0.16 & 17.28 & 2.70\\
HRcam\tablenotemark{a} & 49 & 0.24 & 27.60 & 6.61\\
Herzberg Spectrograph & 34 & 0.24 & 29.11 & 6.86\\
Fabry-Perot & 24 & 0.16 & 19.84 & 3.11\\
Adaptive Optics near-IR imaging & 21 & 0.21 & 20.76 & 4.45\\
SIS\tablenotemark{b} & 18 & 0.12 & 26.82 & 3.24\\
\enddata
\tablenotetext{a}{tip-tilt stablized imager}
\tablenotetext{b}{tip-tilt stablized imager/spectrograph}
\end{deluxetable}

\clearpage

\begin{deluxetable}{lccccccccc}
\rotate
\tablecaption{CFHT's Most Prolific Authors. \label{tbl-3}}
\tablecolumns{10}
\tablehead{
\colhead{} & \multicolumn{5}{c}{Papers} & \colhead{} & \multicolumn{3}{c}{Citations}\\
\cline{2-6} \cline{8-10}
\colhead{Author} & \colhead{Total} & \colhead{\apj} & \colhead{\aap} & \colhead{\aj} & \colhead{\pasp} & \colhead{} & \colhead{Total} & \colhead{per Paper} & \colhead{FCC}}
\startdata
Hutchings, J. & 38 & 6 & 0 & 21 & 8 & & 909  & 23.9 & 1106\\
Davidge, T. & 35 & 12 & 0 & 19 & 1 & & 310 & 8.9 & 492\\
Nieto, J-L. & 16 & 11 & 0 & 1 & 0 & & 282 & 17.6 & 333\\
Le F\`{e}vre, O. & 15 & 3 & 9 & 0 & 0 & & 333 & 22.2 & 525\\
Kormendy, J. & 14 & 12 & 2 & 0 & 0 & & 719 & 51.4 & 1024\\
Boesgaard, A. & 13 & 1 & 0 & 0 & 3 & & 577 & 44.4 & 682\\
Crampton, D. & 13 & 3 & 1 & 7 & 2 & & 200 & 15.4 & 239\\
Richer, H. & 13 & 10 & 1 & 0 & 0 & & 423 & 32.5 & 515\\
Harris, W. & 12 & 3 & 6 & 2 & 0 & & 389 & 32.4 & 513\\
\enddata
\end{deluxetable}


\clearpage 

\begin{deluxetable}{p{6.0cm}p{9.7cm}l}
\rotate
\tablecaption{CFHT's Most Highly Cited Papers. \label{tbl-4}}
\tabletypesize{\scriptsize}
\tablehead{
\colhead{Authors} & \colhead{Title}}
\startdata
\citet{lil95} &  The Canada-France Redshift Survey. VI. Evolution of the
Galaxy Luminosity Function to Z $\approx$ 1\\
\citet{lil91} & A Deep Imaging and Spectroscopic
Survey of Faint  Galaxies\\
\citet{lil96} & The Canada-France
Redshift Survey: The Luminosity  Density and Star Formation History of the
Universe  to Z $\approx$ 1\\
\citet{mcc87} & A Correlation
Between the Radio and Optical Morphologies of Distant 3CR Radio Galaxies\\
\citet{car96} & Galaxy Cluster Virial Masses and Omega\\
\citet{spi82} & Abundance of Lithium in Unevolved Halo Stars and Old
Disk Stars - Interpretation and Consequences\\
\citet{cow96} & New Insight on Galaxy Formation and Evolution From  Keck 
Spectroscopy of the Hawaii Deep Fields\\
\citet{kor85} & Families of Ellipsoidal Stellar Systems and the  Formation of
Dwarf Elliptical Galaxies\\
\citet{tys90} & Detection of Systematic Gravitational
Lens Galaxy Image Alignments - Mapping Dark Matter in Galaxy Clusters\\
\citet{pie94} & The Hubble Constant and Virgo Cluster Distance from
Observations of Cepheid Variables\\
\enddata
\end{deluxetable}

\end{document}